\documentclass[conference]{IEEEtran}
\usepackage[cmex10]{amsmath}
\usepackage{verbatim}
\usepackage[table]{xcolor}
\usepackage{tikz}
\usepackage{gensymb}
\usepackage[ruled,vlined]{algorithm2e}

\usepackage{soul}
\usepackage{multirow}
\usepackage{booktabs,colortbl}
\usepackage{umoline}
\usepackage[utf8]{inputenc}
\usepackage[T1]{fontenc}
\usepackage{ wasysym }
\usepackage{balance}
\usepackage{hhline}

\usepackage{graphicx}
\usepackage{array}
\usepackage{verbatim}
\usepackage{cleveref}
\usepackage{color}
\usepackage{relsize}
\usepackage{tikz}
\usepackage{cite}
\usepackage{epstopdf}
\usepackage{fancyhdr}
\usepackage{graphics}
\usepackage{float}
\usepackage{tablefootnote}
\usepackage{subfigure}

\graphicspath{{./figures/}}

\ifCLASSOPTIONcompsoc

\else

\fi

\ifCLASSINFOpdf

\else

\fi

\begin{document}

\title{False Data Injection Attack Against \\ Power System   Small-Signal Stability}

 \author{%
\IEEEauthorblockN{Mohamadsaleh Jafari, Mohammad A. Rahman, and Sumit Paudyal }\\
\IEEEauthorblockA{
Department of Electrical and Computer Engineering, Florida International University, USA\\
Emails: mjafari@fiu.edu, marahman@fiu.edu,   spaudyal@fiu.edu} \vspace{-25pt}
}


\maketitle



\begin{abstract}

Small-Signal Stability (SSS) is crucial for the control of power grids. However, False Data Injection (FDI) attacks against SSS can impact the grid's stability, hence, the security of SSS needs to be studied. This paper proposes a formal method of synthesizing FDI attack vectors (i.e., a set of measurements to be altered) that can destabilize power systems. We formulate an FDI attack as an optimization problem using AC power flow, SSS model, and stability constraints. The attacker's capability is modeled as the accessibility to a limited set of measurements. The solution of the proposed FDI attack model provides a destabilizing attack vector if exists. We implement the proposed mechanism and evaluate its performance by conducting several case studies using the WSCC 3-machine 9-bus system. 
The case study results showed that the possibility of random FDIs (i.e., with no knowledge of the power system) in launching a destabilizing attack is too low to be successful. However, an intelligent attacker can leverage the grid's knowledge to make the system unstable, even with limited access to the measurements.

\end{abstract}
\vspace{10pt}



\section{Introduction}
\label{Sec:Introduction}

Due to the fast-growing cyber world of complex adaptive systems and the internet, False Data Injection (FDI) attacks are becoming one of the most challenging issues~\cite{song2018intelligent}.
In 2015, there was an FDI attack on the breakers of  Ukrainian power grids that caused a power outage for more than 200 thousand customers for several hours \cite{liang20162015}. FDI attacks can be detrimental to equipment as well. For example, a test carried out by manipulating a diesel generator's circuit breaker demonstrated damage to physical components of the electric power grids \cite{AuroraAttack,salmon2009mitigating}. 
FDI attacks can also lead the system operators to take control actions that may jeopardize the operation of the power grids. FDI attack was first brought up by Liu et al. in the smart grid domain \cite{liu2011false}. Although it may sound common, it particularly means that an attacker stealthy infuses some wrong data into the meter measurements to make the state estimation outputs in the smart grid wrong \cite{2020Springerfalse}. 
This attack is launched in such a way that the bad data detection module of the power grid control center may not detect it.

Stability analysis of power grids is of great importance due to the severe outage problem resulting from losing the stable working region. Small-Signal Stability (SSS) analysis is used to determine the power system's capacity to stay in synchronism when a small disturbance occurs in the system.
Small-signal oscillations in a synchronous generator have been a matter of concern as discussed in several studies \cite{mondal2020power}.
In large interconnected power systems, especially the ones connected through long transmission lines, the SSS issue might threaten the security of the system due to the oscillation among the synchronous generators \cite{rogers2012power}. Eigenvalue analysis methods are widely used for the small-signal analysis of power grids \cite{pai2004small}. Instability occurs when a pair of complex conjugate eigenvalues or a real eigenvalue fall in the right half of the $S$-plane.

We identify and review relevant existing work from literature in two categories: the first focuses on SSS and power flow/Optimal Power Flow (OPF) while the other focuses on FDI attacks in power grids. 
In \cite{adeli2020optimal}, an approach for rescheduling generating units in OPF is proposed which considers SSS constraints of the power system. Generating units are rescheduled if there are any unstable modes in the rotor angles of the generators. 
In \cite{hamon2012stochastic}, a stochastic OPF is proposed with voltage stability and SSS constraints. 
In \cite{li2013eigenvalue}, an eigenvalue optimization-based semi-definite programming model is proposed for SSS constrained OPF. 
A sequential quadratic programming approach combined with gradient sampling for SSS constrained OPF is proposed in \cite{7539295}. 

A redispatching method is presented in \cite{5593195}, which gives the optimal preventive control actions ensuring a given security level of SSS. However, this work does not consider the possibility of launching a stealthy FDI attack.
In \cite{amini2016dynamic}, the authors analyze specific closed-loop, dynamic load altering attacks against power system stability. The analysis follows a mathematical model considering that the attacker controls the changes in the attacked loads based on the feedback from the power system's frequency.
This study considers the DC power flow, which often cannot provide accurate results.
In\cite{zhang2018distributed}, the authors consider the distributed load sharing of autonomous microgrids under FDI attacks, defining the stable region for operating microgrids. However, such analysis at the bulk power level with consideration of synchronous generators is lacking in the literature.  
The authors in \cite{abbaspour2019resilient} propose a resilient control strategy for load frequency control, where they introduce a defense layer capable of detecting and mitigating FDI attacks on power grids.
Brown et al. in \cite{Brown2018} study the possibility of
cyber-physical attacks in creating instability in the power grid; however, the adopted method does not yield an optimal attack vector. 
In \cite{zhou2020cyber}, the impacts of FDI attacks on local/master controllers by compromising communication links are analyzed, and subsequently, a cyber-attack resilient distributed control strategy is proposed in which all the participants can jointly detect and isolate corrupted links. 

To the best of our knowledge, there is no work reported in the literature considering FDI attacks on load measurements and SSS analysis of power systems using an AC OPF framework. In this context, this paper aims at studying the feasibility of launching FDI attacks on the power system against small-signal stability. More importantly, unlike existing works, our formal model can be solved to automatically synthesize potential FDI attack vectors that lead to the system's instability.

\textcolor{black}{
The rest of the paper is organized as follows. In Section \ref{Sec:FormalModeling}, we formulate the SSS model of the power system. In Section~\ref{Sec:FDIAmodeling} we provide the FDI attack as an \textcolor{black}{AC OPF} problem with SSS  as the constraints. In Section~\ref{Sec:CaseStudies}, we provide case studies to assess an FDI attack success considering different levels of the attacker's access to the measurements as well as knowledge of the power system. We conclude the paper in Section~\ref{Sec:Conclusion}.}

\section{Small-Signal Stability Model} 
\label{Sec:FormalModeling}

In this section, we present the formal model of SSS.

\subsection{ Small-Signal Stability of a Dynamic System }
Dynamics  of a power system can be represented by the following generic Differential-Algebraic Equation (DAE), 
\begin{align}
\label{DAE}
    \dot{\textbf{x}} = \textbf{F}_D(\textbf{x,y}) \,, & \\
    \quad
    \textbf{0} = \textbf{F}_A(\textbf{x,y}), &
\end{align}
where $\textbf{x}$ is the state variable vector, and $\textbf{y}$ is the vector of non-state variables.
For small-signal model, linearized form of (\ref{DAE}) is used, which can be written as, 
\begin{equation}
\label{SSSmodel}
    \begin{bmatrix}
        \Delta \dot{\textbf{x}} \\
        \textbf{0}
    \end{bmatrix} = 
    \begin{bmatrix}        \tilde{\textbf{A}}\,&\tilde{\textbf{B}} \\      \tilde{\textbf{C}}\,&\tilde{\textbf{D}}
    \end{bmatrix}
    \begin{bmatrix}
        \Delta \textbf{x} \\
        \Delta \textbf{y}
    \end{bmatrix}. 
\end{equation}
Eliminating $\Delta \textbf{y}$ from (\ref{SSSmodel}), we obtain,
\begin{align}
\label{SSSequ}
    \Delta \dot{\textbf{x}} = \textbf{A} \Delta\, \textbf{x},&\\
    \label{SSSbeginConstr}
\textbf{A}=\tilde{\textbf{A}}-\tilde{\textbf{B}}\, \tilde{\textbf{D}}^{-1} \tilde{\textbf{C}},
\end{align}
where $\textbf{A}$ is  the state matrix. For SSS analysis,  eigenvalues  of   $\textbf{A}$ are computed as, 
\begin{align}
    \label{Eig}
    \boldsymbol{A \phi} = \lambda\, \boldsymbol{\phi},
\end{align}
where $\boldsymbol{\phi}$ is normalized right eigenvector and $\lambda$ is a set of eigenvalues corresponding to $\boldsymbol{\phi}$. For a dynamic system to be small-signal unstable, at least one eigenvalue must lie on positive half-plane, i.e.,
\begin{align}
\label{Eig1}
    \textrm{\bf Real}(\lambda)  > 0.
    \end{align}

\subsection{Power System Small-Signal Model}
Consider $g$ is the set of generator buses including the slack bus, and $l$ is the set of load buses including zero-injection buses. Assume $n (=g \cup l)$ is set of all buses, and  $g'$  represents set of generator buses except the slack bus on the power network. Assume $m$ is alias of $n$. $n_g$ is the number of generators. Let $j=\sqrt{-1}.$ Define complex net power injection, net current injection, and bus voltage vectors as  $\overrightarrow{ \textbf{S}_{n}}={ \textbf{P}_{n}}+j\,{ \textbf{Q}_{n}}$,\,\, $\overrightarrow{ \textbf{I}_{n}}={ \textbf{I}_{n}^r}+j\,{ \textbf{I}_{n}^i}$, and $\overrightarrow{ \textbf{V}_{n}}={ \textbf{V}_{n}^r}+j\,{ \textbf{V}_{n}^i}$. For small-signal analysis purpose, a linearized power grid model around an operating point ($\overrightarrow{\textbf{I}_{g}^0}$, $\overrightarrow{\textbf{V}_{g}^0}$,  $\overrightarrow{\textbf{V}_{l}^0}$) is used, which can be written as, 

\begin{equation}
  \begin{bmatrix}
    \Delta \textbf{I}_g \\
    \textbf{0}
    \end{bmatrix}
    =
    \begin{bmatrix}
    \textbf{Y}_{1} & \textbf{Y}_{2} \\
    \textbf{Y}_{3} & \textbf{Y}_{4}'
    \end{bmatrix}
    \begin{bmatrix}
    \Delta \textbf{V}_g \\
    \Delta \textbf{V}_l
    \end{bmatrix},
    \label{pf1}
\end{equation}
where \textcolor{black}{$\Delta \textbf{I}_{g} = [\Delta\textbf{I}_{1}^{r}, \Delta\textbf{I}_{1}^{i},..,\Delta\textbf{I}_{g}^{r}, \Delta\textbf{I}_{g}^{i}]^T$} is the vector of real and imaginary components of incremental injection currents  $\overrightarrow{\Delta \textbf{I}_{g}}$ at generator buses, $\Delta \textbf{V}_{g}$ is the vector of real and imaginary components of incremental voltage $\overrightarrow{\Delta \textbf{V}_{g}}$ at generator buses, and $\Delta \textbf{V}_{l}$ is the vector of real and imaginary components of incremental voltage $\overrightarrow{\Delta \textbf{V}_{l}}$ at load buses. \textcolor{black}{The sub-matrices $\textbf{Y}_{1}$, $\textbf{Y}_{2}$, and $\textbf{Y}_{3}$  are extracted from the bus admittance matrix $\textbf{Y}$ written as,  
\begin{equation}
\label{NetEqu}
    \textbf{Y}
    =
    \begin{bmatrix}
    \textbf{Y}_{11} & \cdots & \textbf{Y}_{1n} \\
    \vdots & \ddots & \vdots \\
    \textbf{Y}_{n1} & \cdots & \textbf{Y}_{nn} \\
    \end{bmatrix}
\end{equation}
where $\textbf{Y}_{nm} = \Big[\begin{matrix}
  G_{nm} & -B_{nm} \\
  B_{nm} &  G_{nm}
\end{matrix}\Big]$ in which $G_{nm}$ and $B_{nm}$ are the real and imaginary components of the line admittance between buses $n$ and $m$, respectively.
$\textbf{Y}_{4}'$ is obtained by adding load admittance on diagonal of matrix $\textbf{Y}_{4}$, and  $\textbf{Y}_{4}$ is also extracted from $\textbf{Y}$, 
\begin{gather}
\label{modifiedYbus}
     \textbf{Y}_{4_{nn}}' = \begin{bmatrix}
      G_{nn} + G_{L_n} & -B_{nn}+ B_{L_n} \\
      B_{nn} - B_{L_n} &  G_{nn} + G_{L_n}
\end{bmatrix}
\end{gather}
where $G_{L_n}$ and $B_{L_n}$ are the real and imaginary components of the load impedance at the bus $n$.} 

Synchronous generators are modeled using classical representation \cite{kundur1994power}. The DAE that describes generators are  linearized as following,
\begin{align}
\label{MatrixFormOneSGen}
  \Delta \dot{\textbf{x}} &= 
    \textbf{A}_{g}\, \Delta \textbf{x} + \textbf{B}_{g}\, \Delta \textbf{V}_{g}\,, \\
\Delta \textbf{I}_{g} &= 
    \textbf{C}_{g}\, \Delta \textbf{x} + \textbf{D}_{g}\, \Delta \textbf{V}_{g}\,,
\end{align}
where $\Delta \textbf{x} = [\Delta \delta_1 \,\, \Delta \omega_1,\cdots,\Delta \delta_{n_g} \,\, \Delta \omega_{n_g} ]^T$ is the vector of the state variables of power system. The details of matrices $\textbf{A}_{g}$, $\textbf{B}_{g}$, $\textbf{C}_{g}$, and $\textbf{D}_{g}$ are provided in the Appendix and further details in \cite{wang2010modern}. We can obtain the following  constituting matrices of the state matrix $\textbf{A}$
\cite{wang2010modern},  
\begin{align}
 \tilde{\textbf{A}} = \textbf{A}_G ,\ 
        \tilde{\textbf{B}} = [\textbf{B}_G \ \textbf{0}], \
        \tilde{\textbf{C}} = 
        \begin{bmatrix}
            -\textbf{C}_G \\
            \textbf{0}
        \end{bmatrix},
      \nonumber \\       
      \label{SSSendConstr}
        \tilde{\textbf{D}} = 
        \begin{bmatrix}
            \textbf{Y}_{1} - \textbf{D}_G & \textbf{Y}_{2} \\
            \textbf{Y}_{3} & \textbf{Y}_{4}'
        \end{bmatrix}.
\end{align}


\vspace{3pt}
\section{False Data Injection Attack Model}
\label{Sec:FDIAmodeling}

We model the FDI attack as an AC OPF problem with SSS model as the constraints. In this paper we assume that the attacker is able to inject false data into load measurements only. Let us define an attack vector as $\overrightarrow{\mathbf{{S}}_l^{a}}$, which is the  compromise (change) made by the attacker on load information (measurement or forecast). Let's consider an attack of $\overrightarrow{\mathbf{{S}}_l^{a}}$, 
changes the generator setpoints as $\overrightarrow{\mathbf{{A}}_v}$, where $\overrightarrow{\mathbf{{A}}_v}=[\overrightarrow{\mathbf{{P}}_g^a}$ \,\, $\overrightarrow{\mathbf{{V}}_g^a}]^T.$ The FDI attack can be modelled as, 
\begin{align}
\label{eq:OF}
     &\textbf{Min:}  \quad  \, \mathbf {f}( \overrightarrow{\mathbf{{A}}_v})  \\
\label{pfConstrAllbuses}
&\textbf{subject to:}  \nonumber \\ 
    & \overrightarrow{ \textbf{I}_{n}} = \textbf{Y} \,   \overrightarrow{\textbf{V}_n}, 
    \end{align}
    
    \begin{align}
    \label{GPF}
    & \textbf{diag}\left (\overrightarrow{ \textbf{S}_g} +\overrightarrow{ \textbf{S}_g^a}\right)=   \textbf{diag}\left
    (\overrightarrow {\textbf{V}_g}+\overrightarrow{ \textbf{V}_g^a}\right)\,\textbf{diag}\left({\overrightarrow{ \textbf{I}_g}}^*\right),
    \,\,\,\, \\
    \label{LPF}
     &  \textbf{diag}\left (\overrightarrow{ \textbf{S}_l} +\overrightarrow{ \textbf{S}_l^a}\right)=   \textbf{diag}\left
    (\overrightarrow {\textbf{V}_l}\right)\,\textbf{diag}\left({\overrightarrow{ \textbf{I}_l}}^*\right), \,\,\,\, \\
    \label{Vbounds}
     & \underline{\textbf{V}_n}\le |\overrightarrow {\textbf{V}_n}| \le \overline{\textbf{V}_n},\,\,\,\,  \\
     & \textbf{constraints\,\, (5)-(8), (10)-(13)} \nonumber \\
     \label{OPIg}
     & \overrightarrow{\textbf{I}_{g}^0}=\overrightarrow{\textbf{I}_{g}},\,\,\,\,   \\
     & \overrightarrow{\textbf{V}_{g}^0}=\overrightarrow{\textbf{V}_{g}},\,\,\,\,  \\
     \label{OPVl}
     & \overrightarrow{\textbf{V}_{l}^0}=\overrightarrow{\textbf{V}_{l}},
\end{align}
where $\underline{\textbf{V}_n}$ and $\overline{\textbf{V}_n}$ are the minimum/maximum acceptable nodal voltage magnitudes, and $\textbf{diag(\,\,)}$ results in a diagonal matrix. Equation (\ref{pfConstrAllbuses}) represents load flow equations in terms of current injection, Equation (\ref{GPF}) and (\ref{LPF}) represent modeling of generation and load powers, respectively. Equation (\ref{Vbounds}) ensures voltage bounds. Constraints (5)-(8),(10)-(13) are SSS constraints. Constraints (\ref{OPIg})-(\ref{OPVl}) provides an operating point for small-signal model. 

\section{Case Studies}
\label{Sec:CaseStudies}
The proposed method is applied to the WSCC 3-machine 9-bus, which is a test system usually used for small-signal stability analysis. WSCC system has three loads and three generators, as shown in Fig.~\ref{fig:WSCC} and  Table~\ref{tab:loagGenData}.
The nonlinear equations of this study are implemented in Julia for Mathematical Programming (JuMP), which is a domain-specific modeling language for mathematical optimization embedded in Julia \cite{DunningHuchetteLubin2017}. Also, Interior Point Optimizer (Ipopt) is used as the Nonlinear Programming (NLP) solver in these case studies\cite{Ipopt}.
We consider the following case studies:
\begin{figure}[b]
    \centering
    \includegraphics[scale=0.27, keepaspectratio=true]{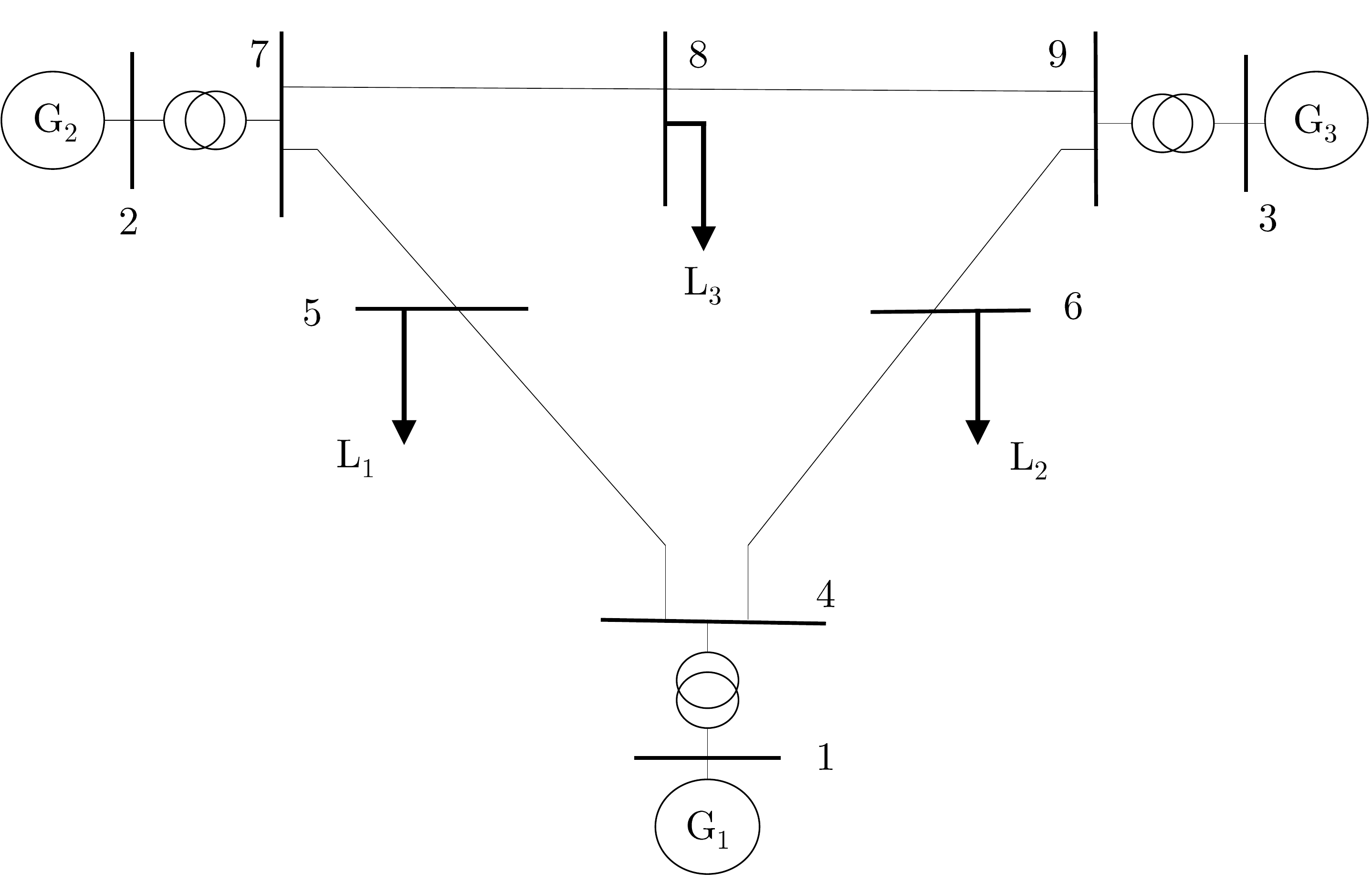}
    \caption{WSCC 9-bus system used for case studies \cite{kundur1994power}.}
    \label{fig:WSCC}
\end{figure}

\begin{table}[t]
\caption{Load data and generator setpoints for WSCC 3-machine 9-bus system \cite{kundur1994power}.}
\small
\vspace{-5pt}
    \centering
    \begin{tabular}{||c|c|c|c|c|c||} 
    \hline
    \multicolumn{3}{|c|}{\textbf{Load}}   &  \multicolumn{3}{|c|}{\textbf{Generator}}  \\ [0.5ex] \hline\hline 
    \textbf{Label} & $\textbf{P}_L$ & $\textbf{Q}_L$ & \textbf{Label} & $\textbf{P}_{sp}$ & $\textbf{V}_{sp}$ \\
    & (MW) & (MVAr) & & (MW) & (pu) \\ \hline
    $L_1$ & 125 & 50  & $G_1$ &  -   & 1.040 \\ \hline
    $L_2$ & 90  & 30  & $G_2$ &  163 & 1.025 \\ \hline
    $L_3$ & 100 & 350 & $G_3$ &  85  & 1.025 \\ 
 \hline
    \end{tabular}
    \label{tab:loagGenData}
    \normalsize
\end{table}

$1)$ \textbf{\textit{Initial Stability Analysis (C1):}} The evaluation of the proposed mechanism starts with implementing a case study ($C1$) to ensure that the power system is stable with the scheduled generation setpoints and loads. As shown in Table~\ref{table:Eigs_C1C2}, eigenvalues are either zero or negative; thus, the system is initially stable.  A destabilizing attack vector 
will be considered as successful if the system becomes unstable after the attack.

\begin{figure}[]
    \centering
    \includegraphics[width=0.47\textwidth]{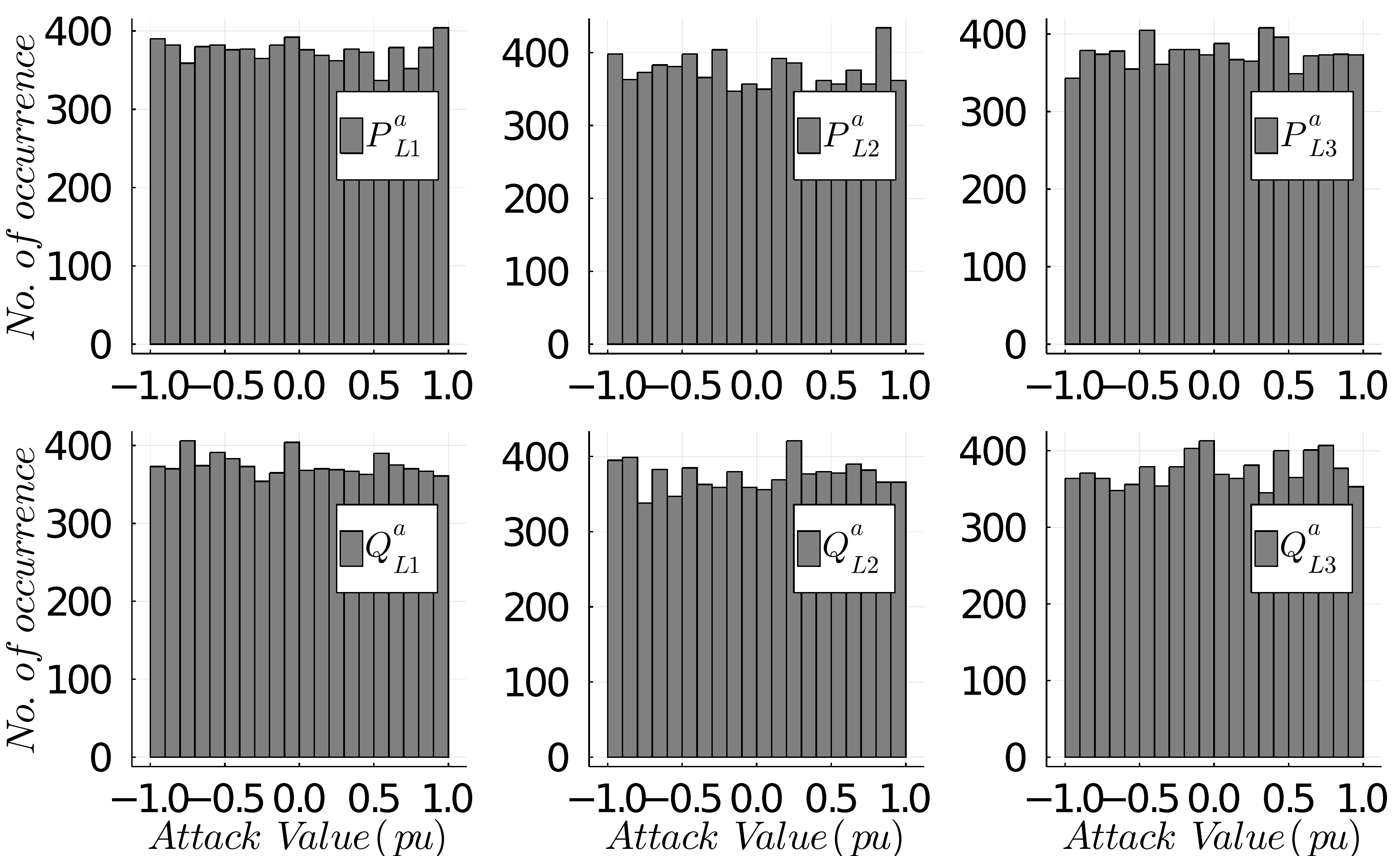}
    \caption{Distribution of random attacks on active and reactive load measurements in \textit{C2}.}
    \label{fig:Hist_Collection}
\end{figure}

\begin{figure}[b!]
\vspace{-6pt}
    \begin{center}
        \subfigure[]{
        \label{fig:groupedbars(DelVG)}
            \includegraphics[width=0.72\columnwidth ]{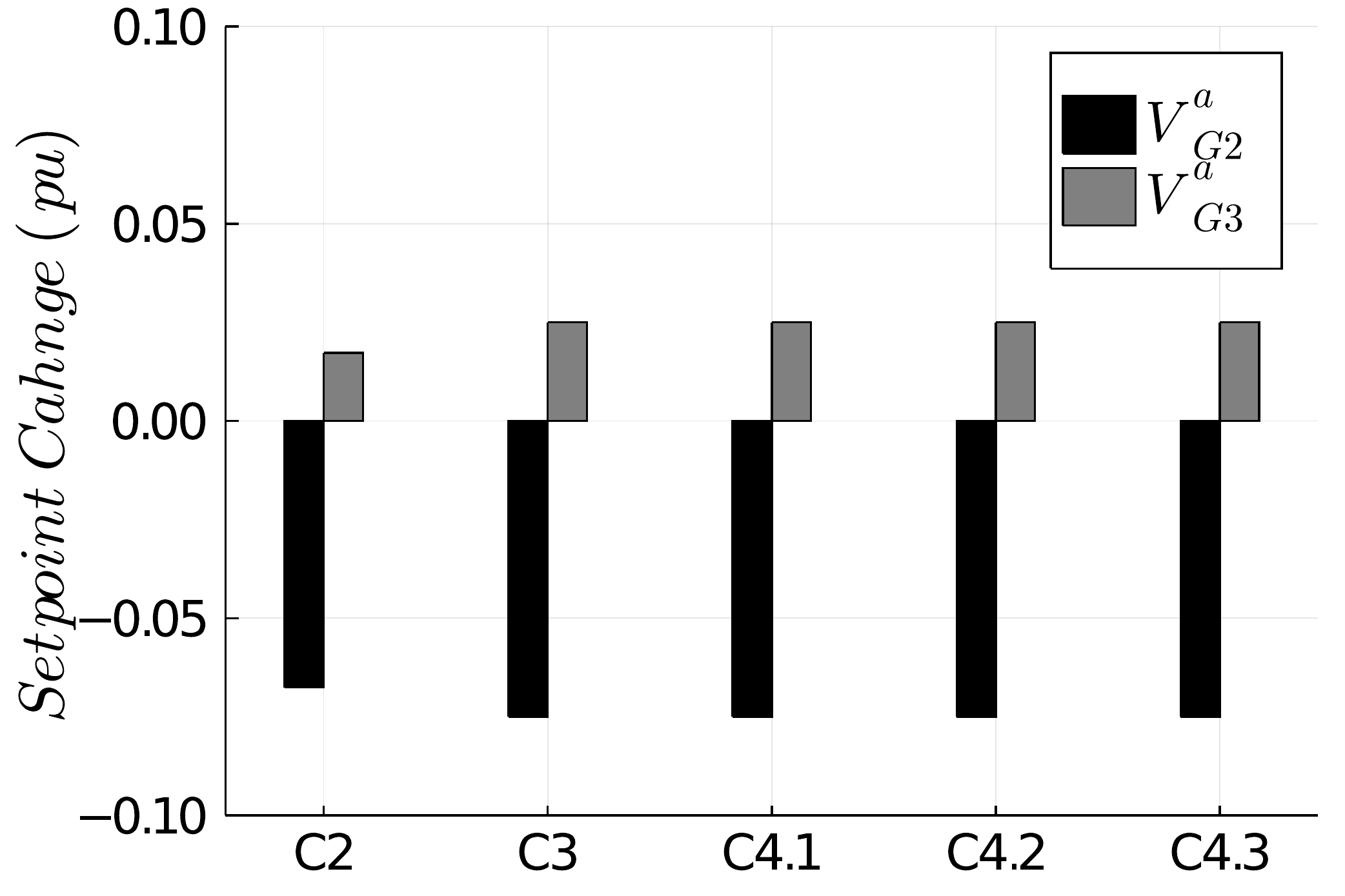}
        }
        \subfigure[]{
        \label{fig:groupedbars(DelPG)}
            \includegraphics[width=0.72\columnwidth ]{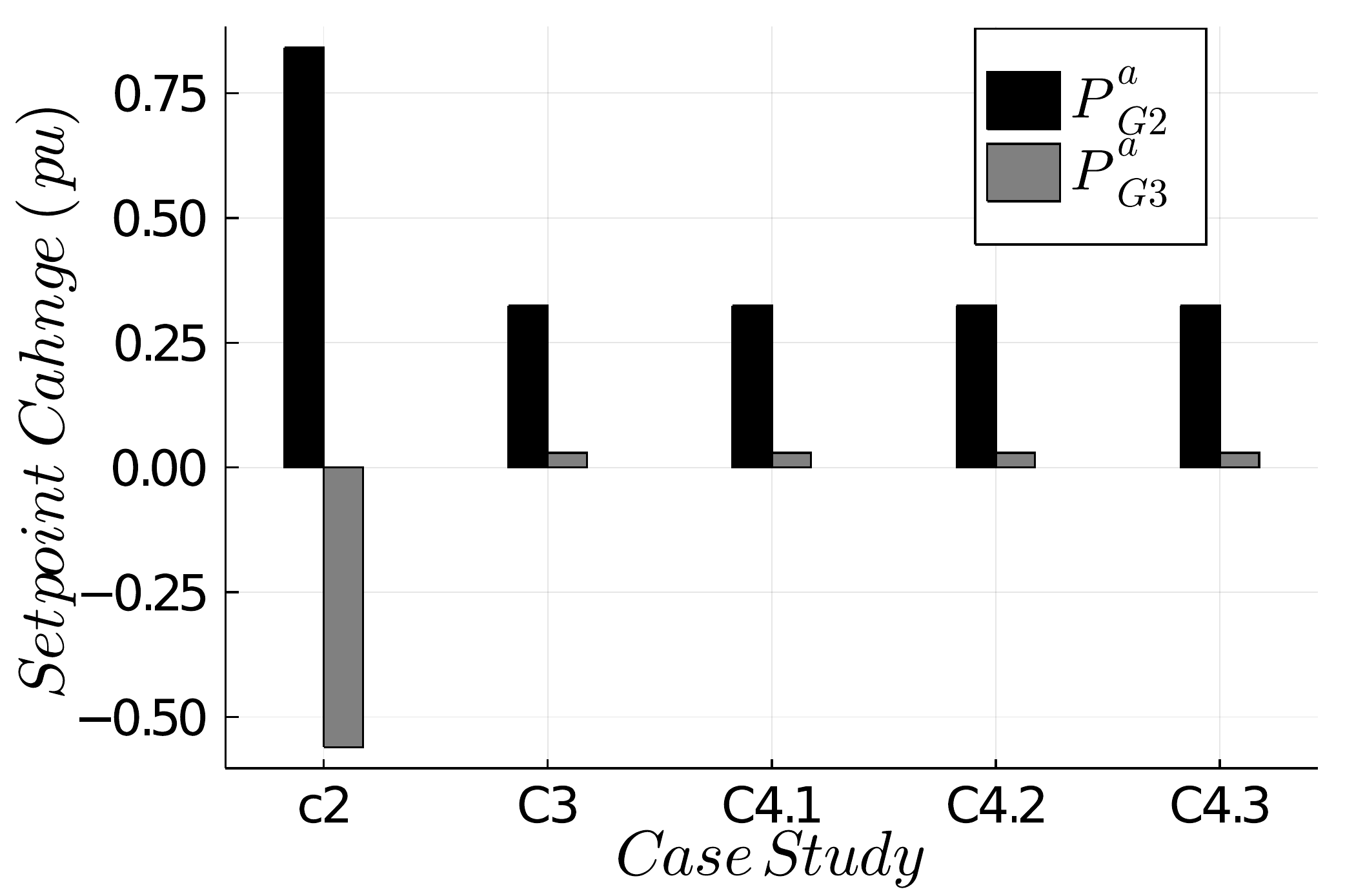}
        }
    \end{center}
    \vspace{-12pt}
     \caption{Changes of generator setpoint values (a) $\textbf{V}_G^a$ and (b) $\textbf{P}_G^a$ due to FDI attacks on load measurements for different case studies.} 
    \label{fig:groupbarsSeparate}
\end{figure}

$2)$ \textbf{\textit{Success Probability of Fully Random Attacks on Load Measurements (C2):}} 
In this case study, we assess the possibility of launching a successful attack if the attacker doesn't have any knowledge about the power system and only has access to load measurements to launch the attack.
In order to implement this scenario, we launch 7,500 uniform random attack vectors of $ \overrightarrow{\mathbf{{S}}_l^{a}}$, and observe the impacts on SSS of the system. The distribution of $ \overrightarrow{\mathbf{{S}}_l^{a}}$ is shown in Fig.~\ref{fig:Hist_Collection}. Among 7,500  uniform random attacks, we obtain only a single successful attack that makes the system unstable. This clearly shows that the success rate of launching a destabilizing attack on the power system is very low ($\approx$ 0.013\%), if the attacker doesn't have enough knowledge about the network and how the power grid operates. Eigenvalues of the successful random attack are listed in Table~\ref{table:Eigs_C1C2}.   Fig.~\ref{fig:groupbarsSeparate} depicts the magnitude of compromise the attacker has to make on $ \overrightarrow{\mathbf{{S}}_l^{a}}$ for a successful attack, which shows for a random successful attack large changes on the load measurements are needed.

\begin{table}[b]
\centering
\caption{Eigenvalues of case studies $C1$ and $C2$.}
\vspace{-3pt}
 \begin{tabular}{||c|c|c||}  \hline
 & Case-$C1$ & Case-$C2$ \\ [0.5ex] \hline
 \multirow{5}{1em}{$\mathbf{\lambda}$}  
 & -0.077 + j0.000      & -3.404 + j0.000 \\
 & -0.074 $\pm$ j5.826  & -0.076 $\pm$ j7.546\\
 & -0.020 $\pm$ j4.247  & -0.005 + j0.000\\
 &  0.000 + j0.000      & 0.000 + j0.000\\ 
 &                      & \textbf{3.296 + j0.000}\\\hline
\end{tabular}
\label{table:Eigs_C1C2}
\end{table}

$3)$ \textbf{\textit{Attack with Full Knowledge and Full Access to Load Measurements (C3):}} In this case study, we assume that the attacker has full knowledge of the power system (network topology, transmission line impedances, and generator parameters) as well as access to all the load measurements to launch an attack. 

$4)$ \textbf{\textit{Attack with Full Knowledge but Limited Access to Load Measurements (C4):}} In this case, we implement the attack assuming that the attacker has limited access to load measurements. We study three distinct scenarios:
\begin{itemize}
    \item  $C4.1$: The attacker can compromise only two of the load measurements ($ L_1\, \text{and}\, L_3 $). 
    \item $C4.2$: The attacker can access only one of the load measurements ($L_1$).
    \item $C4.3$: The attacker can inject false data into the active power measurements only. 
\end{itemize}

\begin{figure}[b!]
\vspace{-6pt}
    \begin{center}
        \subfigure[]
        {
        \label{fig:groupedbars(DelPL)}
            \includegraphics[width=0.71\columnwidth]{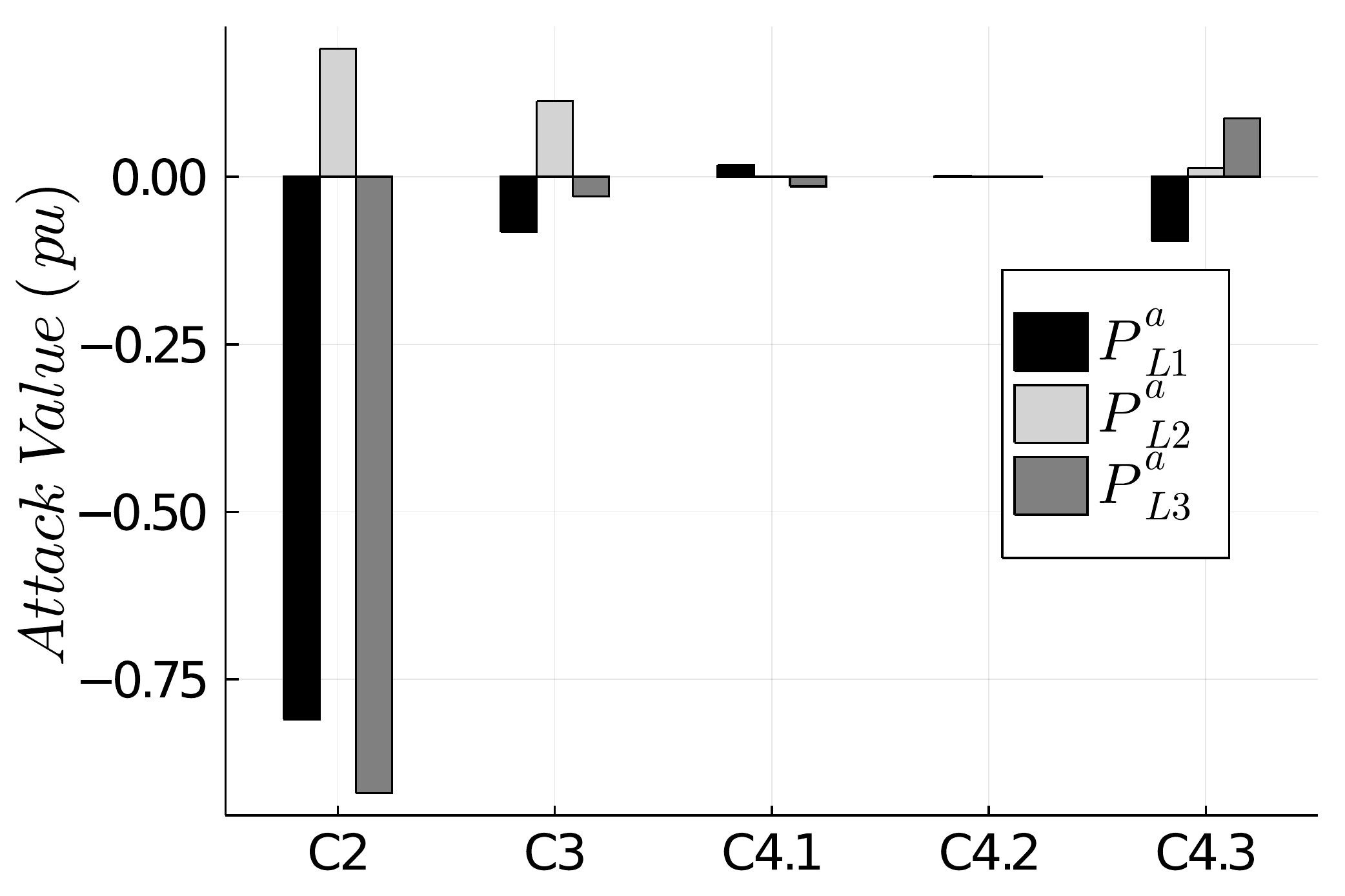}
        }
        \subfigure[]{
        \label{fig:groupedbars(DelQL)}
            \includegraphics[width=0.71\columnwidth ]{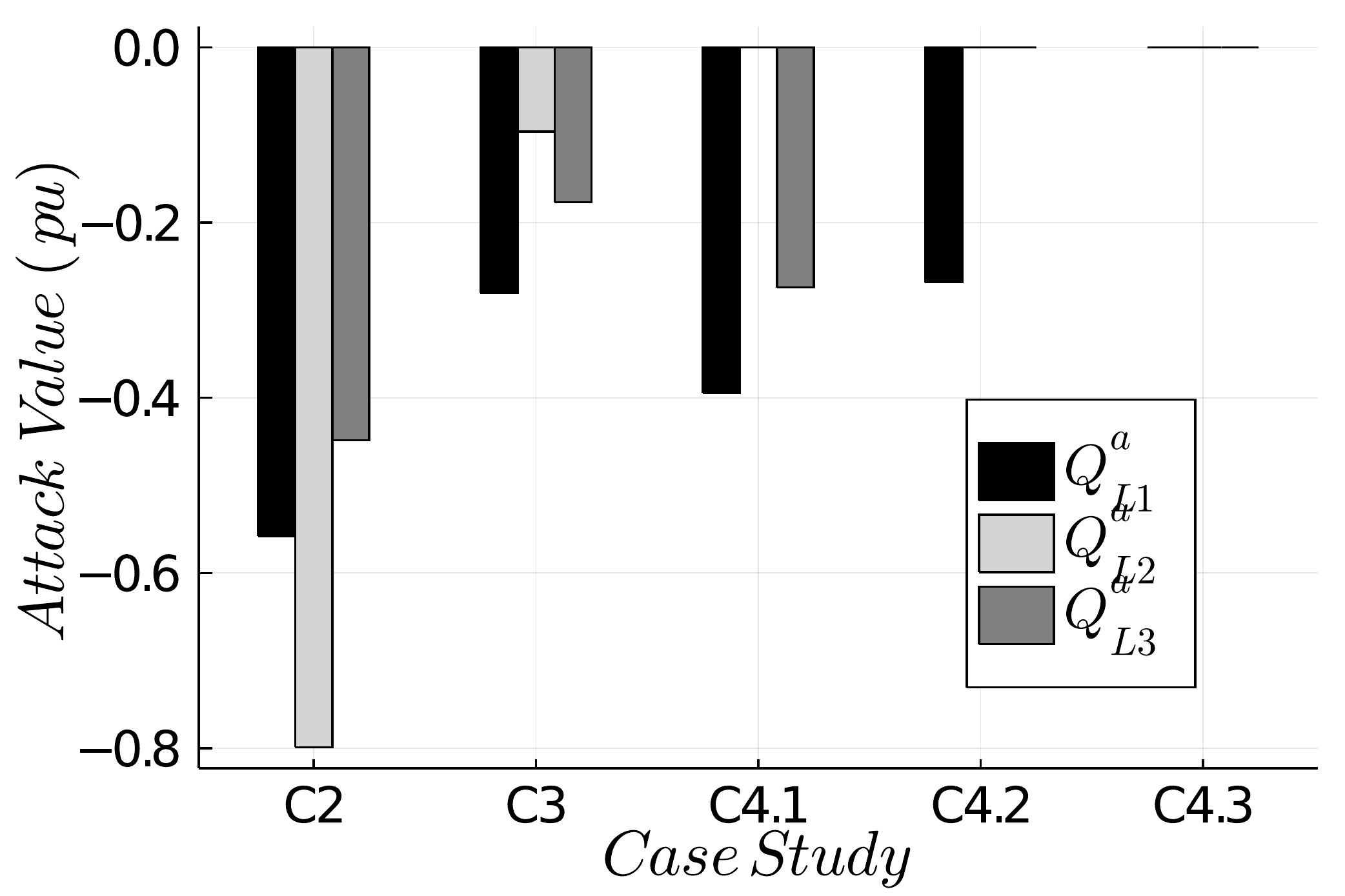}
        }
        \caption{Attack values (a) $\textbf{P}_L^a$ and (b) $\textbf{Q}_L^a$ for different case studies.}
        \end{center}
\end{figure}

Comparing results of $C2$ and $C3$   in \textcolor{black}{Fig}.~\ref{fig:groupedbars(DelPL)}, it is interesting to note that  knowledge of power system can substantially minimize the attack vector to make the system unstable. This is very crucial as the change in the load measurements may not be noticeable to the system operations in many cases, which makes it easier for the attacker to successfully execute the attack. 
Comparing the results, it can also be seen that the attacker may launch a smaller attack on active power loads compared to reactive power; thus, system operators may need to pay attention to expected reactive power loads on the network as well, otherwise, the attack goes stealthy. Based on $C4$, we can observe that attacking only a few load measurements is sufficient for the attacker to make the system unstable if the attacker is knowledgeable about network parameters and power system operation practices.

Additionally, as the results of case studies $C3$ and $C4$ shows, a successful attack vector is larger when more number of measurements are compromised. This is because of the problem's NLP nature and that the solutions obtained are trapped at a local minimum. If the proposed models are able to be convexified, then we expect to see trivial results that a smaller attack would make the system unstable if more number of measurements are attacked. Another possible reason is that since we do not include load attack term $\overrightarrow{\mathbf{{S}}_l^a}$ in the objective function, rather we minimize a function based on the change in generators' setpoints $\overrightarrow{\mathbf{{A}}_v}$ due to a load attack, the AC OPF finds minimal values in terms of generators' setpoint changes. 
\begin{figure}
    \centering
    \includegraphics[width=0.33\textwidth]{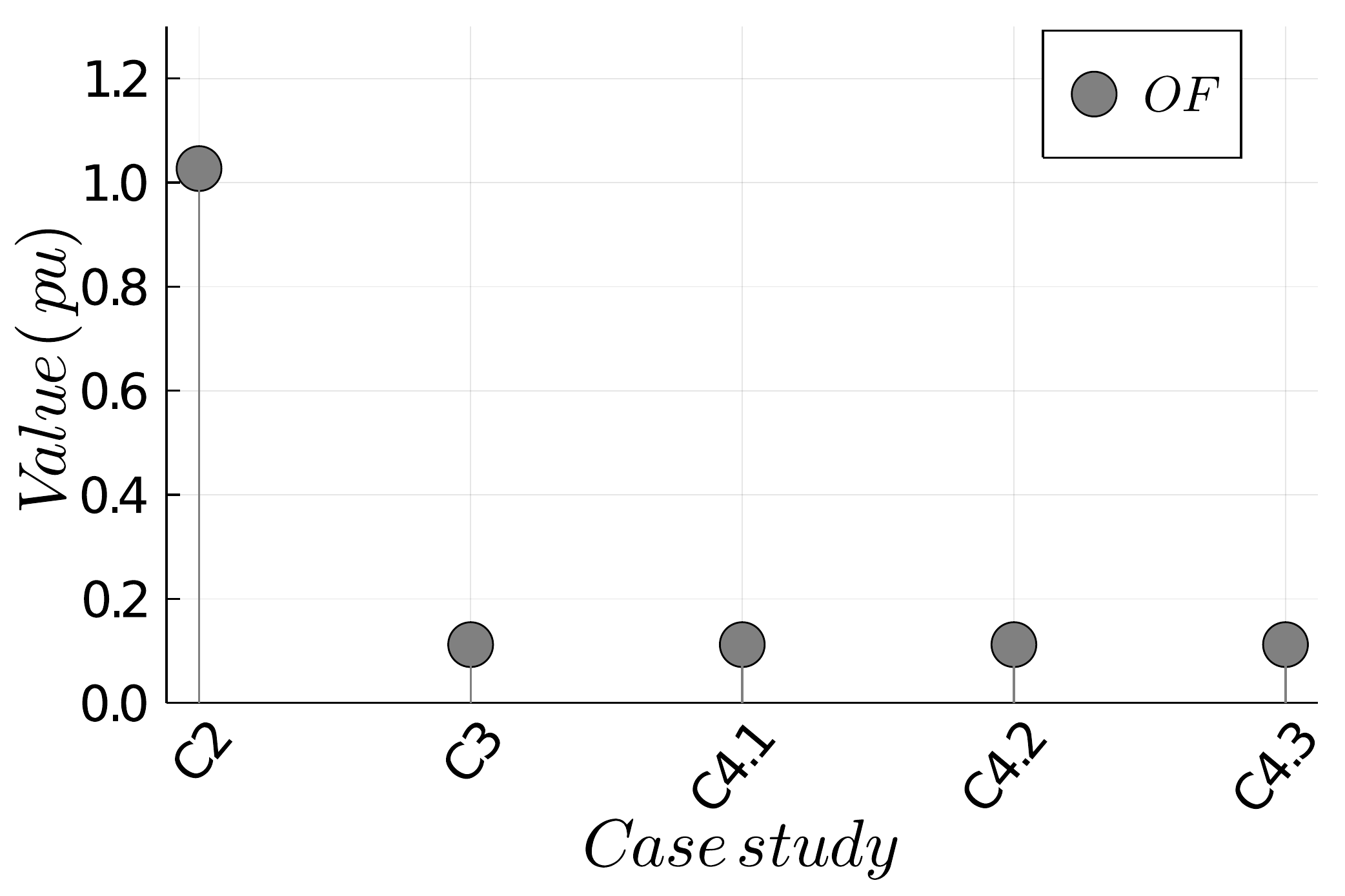}
    \vspace{-9pt}
    \caption{Objective function (OF) values for case studies.}
    \label{fig:OF}
    \vspace{-9pt}
\end{figure}
It can be confirmed from \textcolor{black}{Fig}.~\ref{fig:groupedbars(DelPG)},  Fig.~\ref{fig:groupedbars(DelVG)}, and Fig.~\ref{fig:OF} that the optimizer achieves the same values of $\overrightarrow{\mathbf{{A}}_v}$ and objective functions for cases $C3$ and $C4$. As we mentioned before, these solutions correspond to a local minimum, given the NLP nature of the problem. 

\begin{figure}
    \centering
    \includegraphics[width=0.33\textwidth]{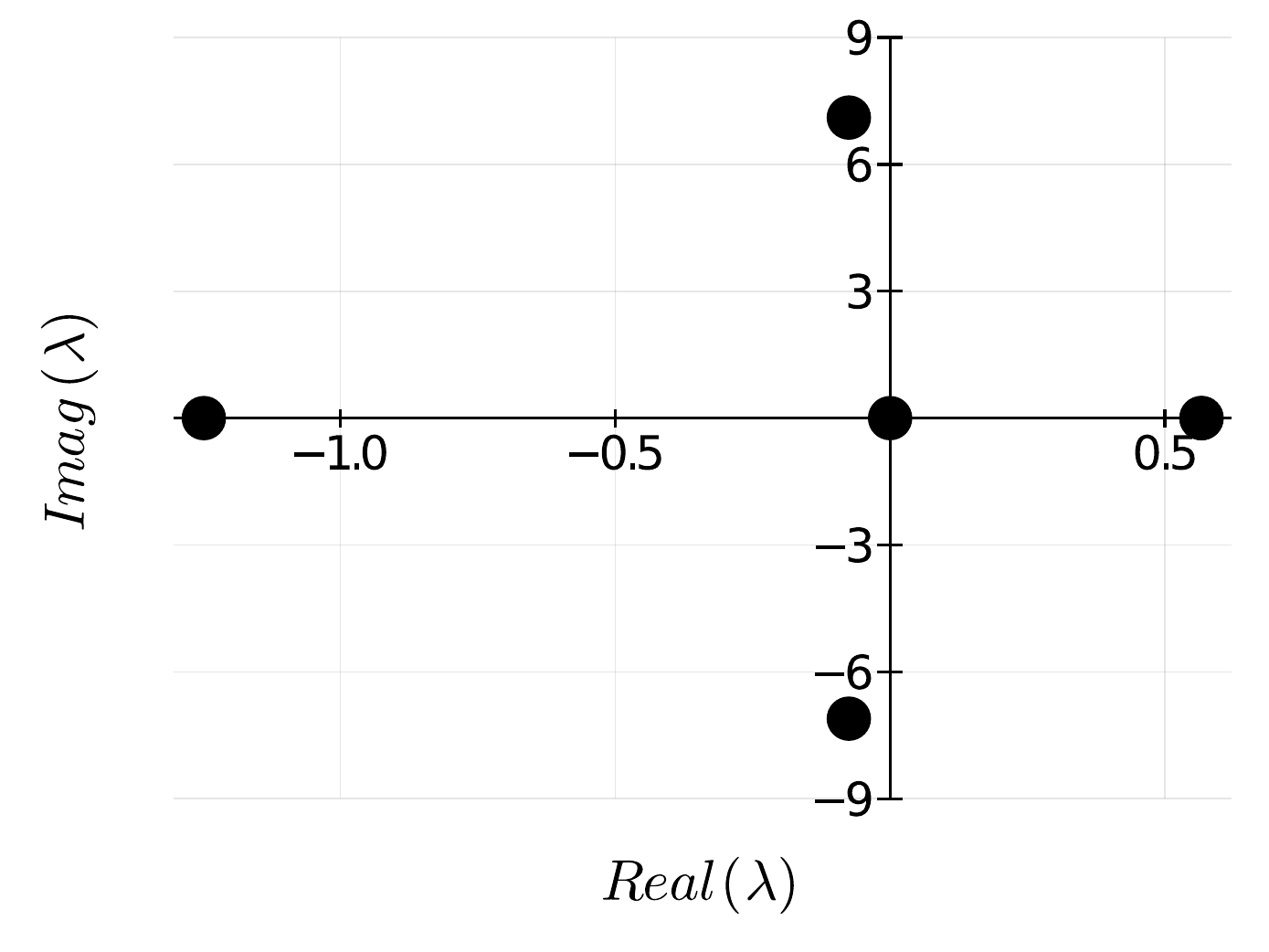}
    \vspace{-9pt}
    \caption{Eigenvalue placement for case studies $C3$ and $C4$.}
    \label{fig:Eigs_Ran+ATO}
    \vspace{-9pt}
\end{figure}

Fig.~\ref{fig:Eigs_Ran+ATO} shows that the eigenvalues for case studies $C3$ and $C4$ are the same. This is also due to the same solution obtained in terms of $\overrightarrow{\mathbf{{A}}_v}$, although the other variables at load buses are different.

\section{Conclusion and Future Work}
\label{Sec:Conclusion}

In this paper, we studied the feasibility of launching a destabilizing FDI attack on the power system in terms of small-signal stability. We used the AC power flow in our study and considered the classical model to implement synchronous generator behaviors in the system. Due to the nonlinearity of the study, we used a Nonlinear Programming (NLP) and optimization solver and proposed a mechanism to find the destabilizing attack vector. 
We conducted several experiments with different levels of attacker's knowledge about the power system was implemented on a 9-bus system and the results showed the proposed mechanism's effectiveness. For example, while the success rate of a random attack is very low ($\approx$ 0.013\%), our mechanism can synthesize minimal successful attack vectors. The future work includes finding a methodology to the feasibility of destabilizing FDI attacks on large scale power with a more detailed order of synchronous generators as well as designing proper defense methods against these attacks.

\vspace{-0.5em}
\bibliographystyle{IEEEtran}
\bibliography{Stability_Attack_References}
%

\vspace{-0.9em}
\section*{Appendix}
\label{Appendix}

\vspace{-12pt}
\begin{align}
    &\textbf{A}_{g_i} = 
    \begin{bmatrix}
        0 & \omega _s \nonumber \\
        0 & \frac{-D_i}{2H_i}
    \end{bmatrix}
\end{align}

\begin{multline*}
    \textbf{B}_{g_i} = 
    \begin{bmatrix}
        0 & 0 \\
        0 & \frac{-E'_i}{2H_i}
    \end{bmatrix}
    \begin{bmatrix}
        -R_{a_i} & X_{q_i} \\
        -X_{q_i} & -R_{a_i} \\
    \end{bmatrix}^{-1}
    \begin{bmatrix}
        \text{sin}\, \delta_{0_i} & -\text{cos}\, \delta_{0_i} \\
        \text{cos}\, \delta_{0_i} & \text{sin}\, \delta_{0_i}
    \end{bmatrix}
\end{multline*}

\begin{multline*}
\textbf{C}_{g_i} = 
    \begin{bmatrix}
        \text{sin}\, \delta_{0_i} & -\text{cos}\, \delta_{0_i} \\
        \text{cos}\, \delta_{0_i} & \text{sin}\, \delta_{0_i}
    \end{bmatrix}^T
    \Big(
     \begin{bmatrix}
        -R_{a_i} & X_{q_i} \\
        -X_{q_i} & -R_{a_i} \\
    \end{bmatrix}^{-1} \\
    \begin{bmatrix}
        V_{q_{0_i}} & 0 \\
       -V_{d_{0_i}} & 0 \\
    \end{bmatrix} - 
    \begin{bmatrix}
        I_{q_{0_i}} & 0 \\
       -I_{d_{0_i}} & 0 \\
    \end{bmatrix} 
    \Big )
\end{multline*}

\begin{multline*}
    \textbf{D}_{g_i} =
    \begin{bmatrix}
        \text{sin}\, \delta_{0_i} & -\text{cos}\, \delta_{0_i} \\
        \text{cos}\, \delta_{0_i} & \text{sin}\, \delta_{0_i}
    \end{bmatrix}^T
    \begin{bmatrix}
        -R_{a_i} & X_{q_i} \\
        -X_{q_i} & -R_{a_i} \\
    \end{bmatrix}^{-1}
    \\
    \begin{bmatrix}
        \text{sin}\, \delta_{0_i} & -\text{cos}\, \delta_{0_i} \\
        \text{cos}\, \delta_{0_i} & \text{sin}\, \delta_{0_i}
    \end{bmatrix}
\end{multline*}
$
\textbf{A}_G = \textbf{diag}\left(\textbf{A}_{g1},\cdots,\textbf{A}_{g_{ng}}\right)$, 
$\textbf{B}_G = \textbf{diag}\left(\textbf{B}_{g1},\cdots,\textbf{B}_{g_{ng}}\right)$, 
$\textbf{C}_G = \textbf{diag}\left(\textbf{C}_{g1},\cdots,\textbf{C}_{g_{ng}}\right)$, 
$\textbf{D}_G = \textbf{diag}\left(\textbf{D}_{g1},\cdots,\textbf{D}_{g_{ng}} \right).$


\end{document}